\documentclass[preprint,showpacs,preprintnumbers,amsmath,amssymb]{revtex4}

\usepackage{graphicx}
\usepackage{dcolumn}
\usepackage{bm}
\usepackage{grffile}
\begin{document}

\title{Counting the number of correlated pairs in a nucleus}

\author{Maarten Vanhalst}
%
\author{Wim Cosyn}
%
\author{Jan Ryckebusch}

\affiliation{Department of Physics and Astronomy,\\
 Ghent University, Proeftuinstraat 86, B-9000 Gent, Belgium}
\date{\today}

\begin{abstract}
We suggest that the number of correlated nucleon pairs in an
arbitrary nucleus can be estimated by counting the number of 
proton-neutron, proton-proton, and neutron-neutron pairs residing in a
relative $S$ state. We present numerical calculations of those amounts 
for the nuclei $^{4}$He, $^{9}$Be, $ ^{12}$C, $ ^{27}$Al, $ ^{40}$Ca,  
$ ^{48}$Ca, $ ^{56}$Fe, $ ^{63}$Cu, $ ^{108}$Ag, and $ ^{197}$Au. The
results are used to predict the values of the ratios of
the per-nucleon electron-nucleus inelastic scattering cross section to
the deuteron in the kinematic regime where correlations dominate. 
\end{abstract}

%


\maketitle

The nucleus is a prototype of a dense quantum liquid with a high
packing fraction \cite{WongNuclearPhys}. 
Naively one could expect some severe medium effects for the
nucleons. Several experimental investigations confirmed the robustness
of the nucleons.  This is for example reflected in the successful use
of the Impulse Approximation (IA) in nuclear reaction theory. In the
IA the bound and free nucleon properties (charges, magnetic moments,
form factors) are considered identical. A few experiments, however,
found indications for medium-modified nuclear properties. Recent
$^{4}$He$(\vec{e},e '\vec{p}) $ measurements
\cite{PhysRevLett.105.072001}, for example, could be described after
implementing medium-modified proton form factors.  Also in comparing
deep inelastic scattering cross sections with those on the deuteron,
one finds that under some kinematics conditions the naive scaling
ratios do not hold. This observation is known as the EMC (European
Muon Collaboration) effect \cite{Geesaman:1995yd} and indicates that
under selected kinematics the whole of the nucleus appears to be more
than the sum of its constituents.

Recently, it was suggested \cite{Weinstein:2010rt} that the magnitude
of the EMC effect can be predicted from the knowledge of the measured
$a_2 (A/D)$ coefficients.  The $a_2 (A/D)$ coefficients are defined as
\begin{equation}
a_2 (A/D) \left( x_{B}, Q ^{2} \right) = \frac {2} {A} \frac 
{\sigma ^{A} \left( x_{B}, Q ^{2} \right)} 
{\sigma ^{D} \left( x_{B}, Q ^{2} \right)} \; , 
\label{eq:pair1}
\end{equation} 
where ${\sigma ^{A} \left( x_{B}, Q ^{2} \right)}$ is the inclusive
$(e,e')$ cross section for the target nucleus $A$ at a particular
four-momentum transfer $Q ^{2}$ and Bjorken $1.4 \le x _{B} = \frac
{Q^{2} } { 2 M \omega} \le 2$ ($M$ is the nucleon mass, and $\omega$
the energy transfer). The observed plateau in the measured $x_{B}$
dependence of $a_2$ for $1.4\le x _{B} \le 2$ is a strong indication
for scattering from a correlated nucleon pair \cite{PhysRevC.48.2451,
  PhysRevLett.96.082501}.  As a matter of fact, the $a_2$ coefficients
can be interpreted as a measure for the effect of short-range
correlations (SRC) in the target nucleus $A$ relative to deuteron 
$D$.  In this paper, we suggest a technique that allows one to
estimate the number of nucleon pairs prone to SRC in an arbitrary
nucleus $A(N,Z)$.  We use these estimates to predict the values of the
coefficients $a_2 (A/D)$.

A time-honored method to quantify the effect of correlations in
classical and quantum systems is the use of correlation functions. The
latter encode those portions of the system that depart from
mean-field behavior.  The realistic (correlated) wave functions 
$\mid \overline{ \Psi}  \rangle $ are constructed by applying a many-body
correlation operator to the mean-field Slater determinant $\mid \Psi
 \rangle$ \cite{Pieper:1992gr, Roth:2010bm} 
\begin{equation}
\mid \overline{ \Psi}\  \rangle =  \frac{1}
{ \sqrt{\langle \ \Psi \mid \widehat{\cal
G}^{\dagger} \widehat{\cal G} \mid \Psi \ \rangle}} \ 
\widehat
{ {\cal G}} \mid  \Psi \ \rangle \; .
\label{eq:coroperator}
\end{equation}
The $\widehat{\cal G}$ reflects the full central, spin and isospin
dependence of the
nucleon-nucleon force but is dominated by the central and tensor
correlations  
\begin{eqnarray}
\widehat{\mathcal{G}}  & \approx & \widehat {{\cal S}}  
\left[ \prod _{i<j=1} ^{A} \left(
1 - g_c(r_{ij})+f_{t\tau}(r_{ij})\widehat{S_{ij}} \vec{\tau}_i
. \vec{\tau}_j 
\right) \right]  \; ,\nonumber \\
& = & \widehat {{\cal S}}  \left[ \prod _{i<j=1} ^{A} \left(
1 - g_c(r_{ij})+ \widehat{t} \left( i,j
  \right)  
\right) \right]  \; ,
\label{eq:pair2}
\end{eqnarray}
where $g_c(r_{12})$ and $f_{t\tau}(r_{12})$ are the central and tensor
correlation function, $\widehat{S_{12}}$ the tensor operator and $
\widehat {{\cal S}} $ the symmetrization operator.  The correlation
functions $g_c$ and $f_{t\tau}$ determine the radial dependence and
magnitude of the correlations. Over the last couple of decades,
various many-body calculations adopting a plethora of techniques
\cite{Pieper:1992gr, Alvioli:2007zz, Bisconti:2006hv, Roth:2010bm}
have made predictions for the correlation functions $g_{c}$ and $f_{t
  \tau}$.  These calculations, confirmed the following robust
features. First, the two-nucleon correlations represent a local
property. This implies that the correlations are universal or only
weakly $A$ dependent \cite{Feldmeier:2011qy}. This means that $g_{c}$
and $f_{t \tau}$ are very much confined to the bulk part of the
nuclear density and only depend on the inter-nucleon distance.  The
universality property implies that the $f_{t \tau} \left( r_{ij}
\right) \widehat{S_{ij}} \vec{\tau}_i . \vec{\tau}_j$ correlation
operator in a nucleus $A$ is not very different from the one that
mixes the $^{3}D_1$ and $^{3}S_1$ wave-function components in
deuterium \cite{Roth:2010bm}.  Second, it was observed that for
moderate relative pair momenta ($300 \le k_{12} \le 600 $~MeV), the
effect of the tensor correlations is dominant \cite{janssen00,
  Schiavilla:2006xx}. As the $\widehat{S_{12}}$ exclusively affects
nucleon pairs in a spin $S=1$ state, it makes the proton-neutron
$(pn)$ correlations to dominate at moderate values of the relative
pair momentum.
%
%
We stress that the universality property does not imply that the
correlation functions $g_c$ and $f_{t \tau}$ are insensitive to model
assumptions.  The correlation functions depend on the choice of the
Hamiltonian, for example. Indeed, a softer Hamiltonian (implying less
correlated wave functions) will require other correlation functions
than a hard Hamiltonian \cite{Anderson:2010aq}.

 Upon computing the response
of the nucleus to some one-body operator $ \widehat{\Omega} = \sum
_{i=1} ^{A} \widehat{\Omega} ^{[1]} (i) $, into lowest order the
effect of the correlations can be implemented by means of an effective
transition operator which includes the effect of the correlations
\cite{Roth:2010bm, janssen00} 
\begin{eqnarray}
& & \widehat{\Omega}^{eff} = \widehat{\cal G}^{\dagger} \
\widehat{\Omega} \  \widehat{\cal G} \approx \widehat{\Omega} 
 +  
\sum_{i<j=1}^A \biggl( \left[ \widehat{\Omega} ^{[1]} (i) + \widehat{\Omega}
  ^{[1]} (j) \right] \nonumber   \\ 
& & \times \left[ - {g}_{c} \left( r_{ij} \right)  + \widehat{t} \left( i,j
  \right) \right] + h.c. \biggr) \; . 
\label{eq:lameff}
\end{eqnarray}
Obviously, through the correlations a typical one-body operator (like
the $\gamma ^{*}$ - nucleus interaction in the IA approximation)
receives two-nucleon contributions which are completely determined by
the product of the correlation functions and the one-body operator.
The nucleon-nucleon correlations are very local and will only affect
nucleon pairs which are ``close''. Accordingly, the correlation
operators $- {g}_c \left( r_{ij} \right)  + \widehat{t} \left( i,j
  \right) $ act as projection operators and will almost exclusively
affect nucleon pairs that reside in a relative $S$ state.

\begin{figure*}
\includegraphics[width=\textwidth]{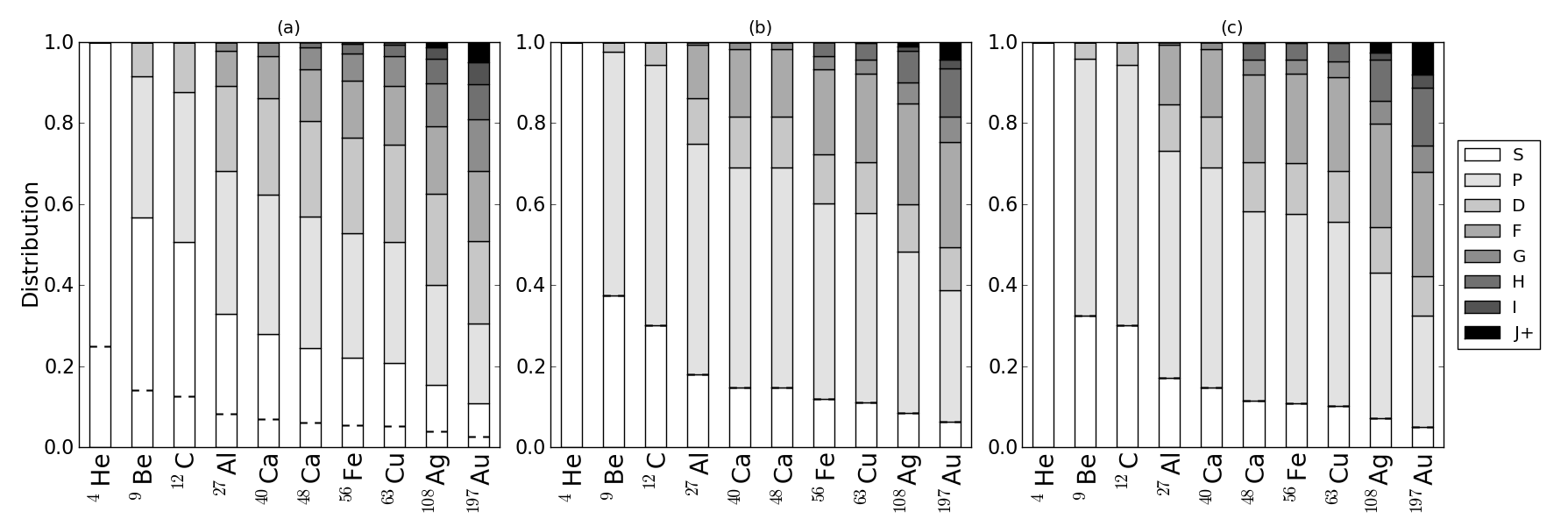} 
 \caption{The distribution of the relative quantum
   numbers $l=S,P,D,F,G,H,I,\ge J$ for (a) the proton-neutron pairs, (b) the proton-proton
   pairs, and (c) the neutron-neutron pairs for the various target
   nuclei. For the proton-neutron pairs there are contributions from
   $^{1}S_{0} (T=1)$ and $^{3}S_{1} (T=0)$. The contribution from the
   $^{1}S_{0} (T=1)$ is indicated by the dashed line. Results are
   obtained in HO basis with $\hbar \omega  \textrm{(MeV)} = 45  A ^{- \frac
     {1}{3}} - 25 A ^{- \frac {2}{3}}$.}
 \label{fig:pairs1}
\end{figure*}

We suggest that the significance of two-nucleons correlations in a
certain nucleus $A(N,Z)$ is proportional to the number of relative $S$
states. In order to compute this number, a coordinate transformation
from $(\vec{r}_1,\vec{r}_2)$ to $ \left(\vec{r}_{12}= \vec{r}_{1} -
\vec{r}_2 , \vec{R} = \frac {\vec{r}_{1} + \vec{r}_{2} } {2} \right)$
is required. The single-particle states in the Slater determinant $
\mid \Psi \ \rangle$ are denoted by $\alpha _{a} = (n_a l_a j_a m_a
t_a)$, where $t_{a} = \pm \frac {1} {2} $ is the isospin quantum
number. In a harmonic-oscillator (HO) basis the normalized and
antisymmetrized two-nucleon wave functions can be written as
\begin{eqnarray}
& & 
\left| \alpha_{a} \alpha _{b} ;  J_R M_R \right> 
  =    
\sum _ {L M_L} \sum _{n l} \sum _{N \Lambda} 
\sum _{S M_S} \sum_{T M_T} 
\frac {1} { \sqrt { 1 + \delta _ {\alpha_{a} \alpha _{b}}}}
\nonumber \\ & & 
\times \left[1-(-1)^{l+S+T}\right] 
\nonumber \\ & & 
\times \; {\cal C} \left (\alpha _{a} \alpha _{b} J_{R} M_{R} ;  (n l N \Lambda) L M_L 
S M_S T M_T \right)
\nonumber \\ & & 
\times \left|(nl,N\Lambda)LM_L,\left(\frac{1}{2} \frac{1}{2}\right)SM_S, 
\left(\frac{1}{2} \frac{1}{2}\right)TM_T \right> 
 \; ,
\label{eq:pair3}
\end{eqnarray}
where $T$ ($S$) is the total isospin (spin) of the pair. Further, 
$\left| n l \right>$
($\left| N \Lambda \right>$) is the relative (center of mass, c.m.) pair wave
function. The explicit expression for the coefficient $ {\cal C}$ can
be found in Eq.~(20) of Ref.~\cite{ryck98a}. With the aid of the above
expression (\ref{eq:pair3}) one can project two-nucleon states in
$(\vec{r}_1,\vec{r}_2)$ on nucleon states in $ \left(\vec{r}_{12},
\vec{R} \right)$ and determine for each pair $(\alpha _a \alpha _b)$
of shell-model states the weight of the various relative ($nl$) and
c.m. ($N \Lambda$) quantum numbers. For two-nucleon states in a non-HO
basis, one can obtain the weights of the various ($nlN\Lambda$)
combinations by expanding the single-particle wave functions in a HO
basis. 


We have computed the $\cal{C}$ coefficients for all target nuclei $A$
either for which the $a_2 (A/D)$ coefficient has been published or for
which one may expect data in the foreseeable future
\cite{Fomin:2008iq}. The Slater determinant is constructed by filling
the single-particle states as they are determined in the nuclear shell
model. We denote the Fermi level for the proton and neutron
single-particle states as $\alpha _F ^p$ and $\alpha _F ^n$. The
quantity
\begin{equation}
\sum _{J_{R} M_{R}} \sum _{\alpha _{a}  \le \alpha _F ^p}
\sum _{\alpha _{b}  \le \alpha _F ^n}
 \left< \alpha_{a} \alpha _{b} ;  J_R M_R \right.
 \left| \alpha_{a} \alpha _{b} ;  J_R M_R \right> = N Z \; ,
\label{eq:pair4}
\end{equation}
determines exactly the number of proton-neutron pairs. Similar
expressions hold for the number of proton-proton $\left( \frac {Z (Z
  -1)} {2} \right)$ and neutron-neutron pairs $\left( \frac {N (N -1)}
{2} \right)$. After inserting the right-hand side of Eq.~(\ref{eq:pair3}) in
the above expression, one can compute how much of each combination
$\left|(nl,N\Lambda) LM_L, SM_S, TM_T \right>$ of pair quantum numbers
contributes to the total number of pairs.  Here, we are particularly
interested in the quantum numbers $(nl)$ of the relative wave
function. We denote the relative orbital angular momentum
$l=0,1,2,\ldots$ as $S,P, D, \ldots$. The numerical calculations get
increasingly more time consuming as $A$ increases due to the
combinatorics of all possible shell-model pairs. The accuracy of the
numerical calculations can be checked against the normalization
condition of Eq.~(\ref{eq:pair4}). In Fig.~\ref{fig:pairs1} we display
the relative contribution of the various $l$ to the pair wave
functions $\left|(nl,N\Lambda)LM_L,\left(\frac{1}{2}
\frac{1}{2}\right)SM_S, \left(\frac{1}{2} \frac{1}{2}\right)TM_T
\right> $ for the nuclei $^{4}$He, $^{9}$Be, $ ^{12}$C, $ ^{27}$Al, $
^{40}$Ca, $ ^{48}$Ca, $ ^{56}$Fe, $ ^{63}$Cu, $ ^{108}$Ag, and $
^{197}$Au.  It is obvious that with increasing $A$ a smaller fraction
of the nucleon pairs resides in a relative $S$ state.  Whereas, for
$^{12}$C about 50\% of the $pn$ pairs has $l=0$ for the heaviest
nucleus $^{197}$Au this is a mere 10\%.  Accordingly, with increasing
$A$, a smaller and smaller fraction of the nucleon-nucleon pairs will
be prone to correlation effects.  In addition, there is a strong
isospin dependence as the fraction of the proton-neutron pairs
residing in a relative $S$ state is substantially larger than for
proton-proton and neutron-neutron pairs. 

\begin{figure}
 \centering
 \includegraphics[width=0.9\textwidth]{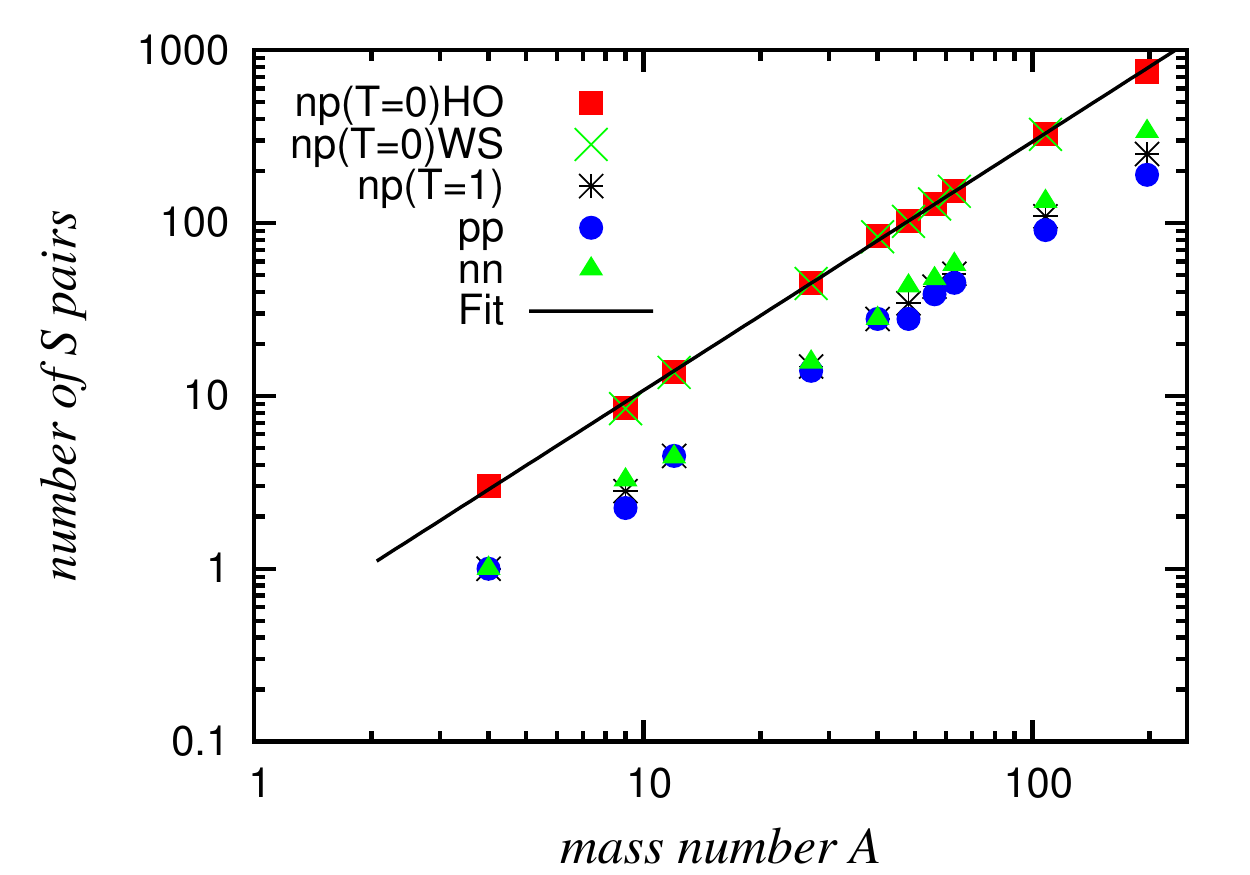}
 \caption{(Color online) The computed number of $pp$, $nn$ and $pn$
   pairs with $l=0$. For $pn$ we discriminate between $^{3}S_1(T=0)$ and
   $^{1}S_0(T=1)$. Unless indicated otherwise the results are for a HO
 basis. For the $^{3}S_1(T=0)$ $pn$ pairs also the predictions in a WS
 basis are shown. The parametrizations for the WS potentials are from
 Ref.~\cite{2007arXiv0709.3525S}.}
 \label{fig:pairs3}
\end{figure}

\begin{figure}
 \centering
 \includegraphics[width=\textwidth]{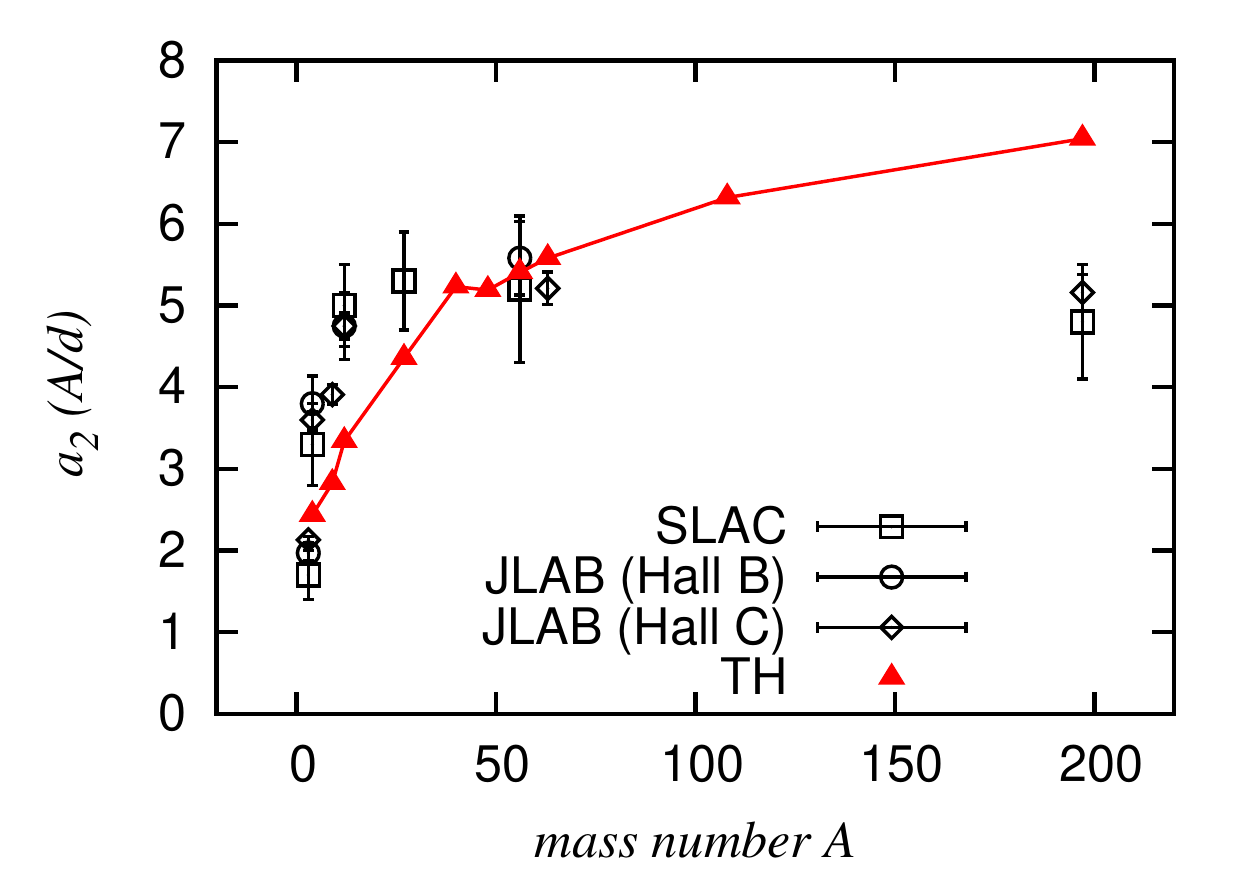}
 \caption{(Color online) The computed values for the $a_2 (A/D)$ for
   various nuclei. The data are from Refs.~\cite{PhysRevC.48.2451}
   (SLAC), \cite{PhysRevLett.96.082501} (JLAB Hall~B) and 
\cite{Fomin:2011ng} (JLAB Hall~C). The triangles denote the
   theoretical predictions obtained with the
   Eq.~(\ref{eq:pairfinal}).}
 \label{fig:pairs4}
\end{figure}

\begin{table}
\begin{center}
\begin{tabular}{c|cccccc}
\hline \hline
A 		& Ref.~\cite{PhysRevC.48.2451} 		& Ref.~\cite{PhysRevLett.96.082501}		& Ref.~\cite{Weinstein:2010rt}  & Ref.~\cite{Fomin:2011ng}		& Eq.~(\ref{eq:pairfinal}) 	\\
\hline
$^4$He 		& $3.3 \pm 0.5$	& $3.80 \pm 0.34$	& 			& $3.60 \pm 0.10$ 	& 2.4 	\\
$^{9}$Be   	& 		& 		 	& $4.08 \pm 0.60$ 	& $3.91 \pm 0.12$ 	& 2.8	\\
$^{12}$C   	& $5.0 \pm 0.5$	& $4.75 \pm 0.41$	& 			& $4.75 \pm 0.16$ 	& 3.3	\\
$^{27}$Al 	& $5.3 \pm 0.6$	&  			& $5.13 \pm 0.55$	& 			& 4.4	\\
$^{40}$Ca  	& 		& 		 	& $5.44 \pm 0.70$	& 			& 5.2	\\
$^{48}$Ca  	& 		& 		 	& 			& 			& 5.2	\\
$^{56}$Fe  	& $5.2 \pm 0.9$	& $5.58 \pm 0.45$	& 			& 			& 5.4	\\
$^{63}$Cu  	& 		& 		 	& 			& $5.21 \pm 0.20$ 	& 5.6	\\
$^{108}$Ag 	& 		&  		 	& $7.29 \pm 0.83$ 	& 			& 6.3	\\
$^{197}$Au 	& $4.8 \pm 0.7$	& 		 	& $6.19 \pm 0.65$ 	& $5.16 \pm 0.22$	& 7.0	\\ 
\hline \hline 
\end{tabular}
\end{center}
\caption{The $a_2 (A/D)$ values for
  various nuclei. The data from direct measurements of the nucleus to deuteron 
  cross sections are from Refs.~\cite{PhysRevC.48.2451}
  (SLAC), \cite{PhysRevLett.96.082501} (JLAB Hall B) and 
  \cite{Fomin:2011ng} (JLAB Hall C). The values of Ref.~  \cite{Weinstein:2010rt} are phenomenological extractions based on the measured EMC data
  and the observed linear correlation between the magnitude of the EMC
  effect and the measured $a_2$ scaling factor. The quoted values of Ref.~\cite{Fomin:2011ng} are the raw ratios. Ref.~\cite{Fomin:2011ng} also contains corrected values for $a_2$ which are about 15\% smaller. 
}
 \label{tab:pairs1}
\end{table}

Naively, one could expect that the number of correlated $pn$ ($pp$)
pairs in a nucleus scales like $NZ$ ($\frac {Z(Z-1)} {2}) \sim A
^{2}$. As illustrated in Fig.~\ref{fig:pairs3} our calculations rather
indicate that the number of pairs that are prone to correlation
effects follows a power law $\sim A^{1.44 \pm 0.01}$. As a matter of
fact, we find that the power law is very robust. Calculations with
Woods-Saxon (WS) wave functions, for example, result in a computed
number of $S$ states that is very close (order of one percent) to the
HO predictions. The $N-Z$ asymmetry is reflected in an unequal number 
of $pp$, $nn$, and $pn$ $^{1}S_0(T=1)$ pairs. We stress that the ratio
of the $nn$ to $pp$ $^{1}S_0(T=1)$ pairs can be considerably smaller
than predicted by naive $\frac{N (N-1)} {Z(Z-1)}$ combinatorics. For
Au, for example, one expects a ratio of 2.24 whereas the data of
Fig.~\ref{fig:pairs3} lead to 1.77.

Now, we wish to connect the number of pairs with $l=0$ with the
measured values of $a_2 (A/D)$. In an inclusive $A(e,e')$ process the
correlated part of the electron-nucleus ($eA$) response (corresponding
with the last two terms in Eq.~(\ref{eq:lameff})) can be probed by
selecting events $1.4 \le x _{B} \le
2$. The magnitude of the response is proportional with a product of
two terms. First, the number of pairs that are prone to SRC, and, second, the
value of the correlation functions evaluated at the relative momentum
of the pair. Indeed, as is pointed out in
Refs.~\cite{Frankfurt:2008zv, Ryckebusch:1996wc} in the kinematical 
regime where correlations are probed, the $eA$ response
obeys $\sim F(P) \sigma _{eNN} (k_{12})$, where $P$ is the
c.m. momentum of the correlated pair on which the absorption takes
place and $F(P)$ is the corresponding c.m.~distribution (the
combination $F(P) \sigma _{eNN} (k_{12}) $ is referred to as the decay
function in Ref.~\cite{Frankfurt:2008zv}). The $\sigma
_{eNN}$ stands for the elementary cross section for electron
scattering from a correlated $NN$ pair. The $\sigma _{eNN}$ contains
the Fourier-transformed correlation functions $g_c(k_{12})$ and
$f_{t\tau}(k_{12}) $ evaluated at the relative momentum $k_{12}$ of
the pair. An analytic expression for $\sigma _{epp}$ can be found in
Ref.~\cite{Ryckebusch:1996wc}. It is worth stressing that given the
kinematics, there are two possible values of $k_{12}$ corresponding
with photoabsorption on nucleon ``1'' and photoabsorption on nucleon
``2'' of the pair. The dominant contribution to the inclusive
$A(e,e')$ cross section for $1.4 \lesssim x_{B} \lesssim 2$ stems from
pairs with $k_{F} \lesssim k_{12} \lesssim 2 k_{F}$, with $k_F$ the
Fermi momentum. In that momentum region, the $g_c(k_{12})$ is
substantially smaller than $f_{t\tau}(k_{12}) $, which causes the
tensor correlated $pn$ pairs to dominate \cite{Piasetzky:2006ai}
\cite{Subedi:2008zz} \cite{PhysRevLett.105.222501}. 

The universality of the tensor correlations, which translates to the
weak $A$ dependence of $f_{t\tau}(k_{12}) $, allows one to assume that
the cross section $\sigma _{epn}$ for electron scattering from a
correlated proton-neutron pair in the nucleus will almost equal the
one for electron scattering from the deuteron, provided that the cross
sections are evaluated at equal values of the high relative momentum
$k_{12}$ of the pair. In a symbolic way, this feature can be expressed throught the scaling relation $\sigma _{epn}
(k_{12}) \approx \sigma_{eD} ( k_{12}) $. This property is related to
the fact that at high momenta the nuclear momentum distributions
$n^{A}(k)$ are very much like scaled deuteron momentum distributions:
$n^{A}(k) \approx C^{A} n^{D} (k)$, where $C^{A}$ is a measure for the
amount of $pn$ correlations in $A$ \cite{Feldmeier:2011qy} \cite{CiofidegliAtti:2009qc}.  

With the above-mentioned scaling relation $\sigma _{epn} (k_{12})
\approx \sigma_{eD} ( k_{12})$ valid at high relative momenta, one can
transform the ratio of of Eq.~(\ref{eq:pair1}) (the per-nucleon
electron-nucleus inelastic scattering cross section to the deuteron)
into the form
\begin{eqnarray}
a_2 (A/D) & = & \frac {2} {A} \frac { \int _{PS} d \vec{k}_{12}  d
  \vec{P} F(P) 
B_{l=0} ^{np} (A) \sigma _{epn}
(k_{12})  }
{ \int _{PS} d \vec{k}_{12} \sigma_{eD} (k_{12}) } \; ,
\nonumber \\
& \approx & \frac {2} {A} B_{l=0} ^{np} (A) \int _{PS}   d \vec{P}
F(P) \; , 
\label{eq:pairfinal}
\end{eqnarray}  
where the integrations extend over those parts of the phase space (PS)
which are compatible with $1.4\le x_{B} \le 2 $.  The quantity
$B_{l=0} ^{np}(A)$ is the number of $pn$ pairs in a relative $\left|
  n, l \right>$ state with the quantum numbers of the deuteron,
$^{3}S_{1} (T=0)$. One can estimate the $B_{l=0} ^{np}(A)$ from
Eq.~(\ref{eq:pair3}) by combining the computed coefficients for all
possible $\left|(n =0 \;, l=0, N\Lambda)LM_L, S=1 M_S, T=0 M_T=0
\right> $ combinations. In Fig.~\ref{fig:pairs3} we have summed over
all possible $n$ to obtain the total amount of $l=0$ states. For all
target nuclei, the $n=0$ contribution dominates, but its relative
importance decreases with growing $A$. The $n=0$ represents 100\% of
the $l=0$ $pn$ states for $^4$He, about 80\% for the medium-heavy
nuclei (Ca, Fe, Cu), 70\% for $^{108}$Ag, and 62\% for $^{197}$Au.
Pairs residing in a $\left| n \ne 0, l =0 \right>$ state have a much
smaller chance of being ``close'' than their $\left| n = 0, l =0
\right>$ counterparts and are less prone to SRC effects. We assume
that only $\left| n = 0, l =0 \right>$ proton-neutron pairs contribute
to $B_{l=0} ^{np} (A)$.

The c.m.~motion of the pair in finite nuclei (absent in the deuteron)
and the imposed conditions in $x_{B}$ make that a fraction of the
correlated proton-neutron pairs are not counted in the $A(e,e')$
signal in the numerator of Eq.~(\ref{eq:pair1}). The $F(P)$ for a
nucleus $A$ can be reliably computed in a mean-field model. Indeed,
the $^{12}$C$(e,e'pp)$ measurements of Ref.~\cite{Blomqvist:1998gq}
determined $F(P)$ over a large $P$ range and observed it to be
compatible with a mean-field prediction.  We have performed
Monte-Carlo simulations in order to determine the correction factor $
\int _{PS} d \vec{P} F(P)$ for all nuclei which are considered
here. We find that for $A> 4$ about 25\% of the correlated pairs are
excluded from the experimentally scanned phase space due to the
c.m.~motion of the correlated pair. From the simulations we observed
that the correction factor is only slightly mass number dependent.
With this correction factor, the Eq.~(\ref{eq:pairfinal}) allows us to
make predictions for the $a_2 \left( A /D \right)$. The predictions
are contained in Fig.~\ref{fig:pairs4} and Table~\ref{tab:pairs1} and
compared with experimental data.  One striking observation from our
calculations, is that the predicted $a_{2} \left( A /D \right) $ for
$^{40}$Ca and $^{48}$Ca are identical and equal to 5.2. On the basis
of naive $NZ$ combinatorics one may have expected a 30\% difference
between the two.  For heavier target nuclei, the data seem to suggest
that the $a_{2} \left( A /D \right) $ coefficient saturates. Our
calculations predict a strong linear rise in the $A$ dependence of the
$a_2$ for $A \lesssim 40$.  At higher $A$ one enters a second regime
with a much softer linear rise with $A$. Our calculations increase
linearly with $\log(A)$ and tend to underestimate the data at low $A$
and overestimate the data for the heavier nuclei. Final state
interactions, for example, which are neglected in this work, may
induce some additional $A$ dependence in the $a_2$ ratio
\cite{Fomin:2011ng}.  It is clear that more data are needed to
establish the situation at large $A$. The observed phenomenological
linear relationship between the scaling factor $a_2$ and the magnitude
of the EMC effect \cite{Weinstein:2010rt} gives $a_{2} = 7.29 \pm 0.83
$ for Ag and $a_{2} = 6.19 \pm 0.65 $ for Au, values that are not
inconsistent with our results.

In conclusion, we suggest that the number of correlated pairs in a
nucleus is proportional with the number of relative $S$ two-nucleon
states.  We find this number to obey a power law $d A^{1.44\pm 0.01}$
with $d=0.39 \pm 0.02$ ($d=0.13 \pm 0.01$) for $T=0$ ($T=1$)
proton-neutron pairs.  The power law is robust in that it is
independent of the choices made with regard to the single-particle
wave functions.  We have used the computed amount of $T=0$ $pn$ pairs
to predict the value of the measured $a_{2} (A/D)$ coefficients, which
provide a measure of the number of correlated pairs in the target
nucleus $A$ relative to the deuteron.  The observed power law in the
number of relative $S$ states translates to a linear increase of
$a_{2}$ with $\log(A)$. We observe that our predictions
are not inconsistent with the trend and magnitude of the data, lending
support to our suggestion.

This work was supported by the Fund for Scientific Research
Flanders. We are grateful to Kris Heyde for useful discussions and
comments.

\bibliography{./myreferences2.bib}

\end{document}